\documentclass[10pt,sigconf,letterpaper,nonacm]{acmart}

\usepackage{amsmath,amsfonts,bbm}
\usepackage{tabularx, array}
\usepackage{caption, subcaption}
\usepackage{url}
\usepackage{algorithmic}
\usepackage{graphicx}
\usepackage{textcomp}
\usepackage{xcolor}
\usepackage{xspace}

\usepackage{tikz}
\usetikzlibrary{external}
\usepackage{pgfplotstable}
\pgfplotsset{compat=1.18}

\AtBeginDocument{%
  }

\begin{document}

\newcommand{\name}{LINC\xspace}

\newcommand{\showComments}{yes}
\newcommand{\note}[2]{%
    \ifthenelse{\equal{\showComments}{yes}}{ \textcolor{#1}{#2}}{}%
}
\newcommand{\manya}[1]{\note{blue}{[Manya: #1]}}
\newcommand{\muriel}[1]{\note{red}{[Muriel: #1]}}
\newcommand{\benoit}[1]{\note{orange}{[Benoit: #1]}}

\newcommand{\uncoded}{Uncoded\xspace}
\newcommand{\inr}{In-Network Retransmission\xspace}
\newcommand{\ete}{End-to-End Coding\xspace}
\newcommand{\INR}{INR\xspace}

\newcommand{\figWidth}{4cm}
\newcommand{\figHeight}{4cm}
\newcommand{\graphFontSize}{\footnotesize}

\title[\name]{\name: An In-Network Coding Approach to Tame Packet Loss in Hybrid Wireless-Fiber Backbones}

\author{Benoit Pit-Claudel, Muriel Médard, Manya Ghobadi}

\affiliation{%
  \institution{MIT}
  \city{Cambridge}
   \state{Massachusetts}
   \country{USA}
}
 \renewcommand{\shortauthors}{Pit-Claudel et al.}

\begin{abstract}
The emergence of ultra-low latency applications, such as financial transactions, has driven the development of hybrid backbone networks that rely on fiber, satellite, and microwave links (Figure~\ref{fig:illustration}). Despite providing low latencies, these hybrid networks suffer from occasional environmental packet loss caused by poor weather, construction, and line of sight blockage.
Paradoxically, today's hybrid backbones rely on conventional transport protocols that take packet loss to signal network congestion, as opposed to transient environmental obstacles. A common approach to address this challenge is to use network coding (NC) between the end hosts to recover from these occasional packet loss events. However, current NC proposals assume full access to the end-hosts' stack to perform end-to-end encoding/decoding operations.
In this paper, we introduce \name, a novel system that provides in-network NC capabilities to mitigate environmental packet loss events without requiring cooperation from the end hosts. \name uses a systematic block coding approach on a link-by-link basis, encoding and decoding packets inside the network. We model the tradeoff in goodput between end-to-end retransmissions and redundant packets introduced by \name, and propose an optimization formulation to determine the optimal choice of coding parameters.
Our simulations on real-world backbone topologies demonstrate that \name reduces the end-to-end latency by up to $18\%$ by eliminating unnecessary retransmissions.
\end{abstract}

\maketitle

\section{Introduction}\label{sec:introduction}

End-to-End (E2E) latency and goodput play a critical role in Quality of Service (QoS) metrics of ultra-low latency applications. Major internet service providers reported a direct correlation between latency increases and business revenue~\cite{googleLatency,Pudica}; for example, AWS reported in~\cite{amazonLatency} a $1\%$ loss in sales per $100$~ms of added latency. Conventional backbone networks heavily rely on fiber-optic deployments to carry both throughput and latency-sensitive applications.

However, fiber deployments depend on geographical constraints: laying fiber in some areas can be impossible or prohibitively expensive resulting in long detours and added latency. As a result, emerging ultra-latency-sensitive application providers deploy hybrid wireless-fiber backbones to carry traffic through a mix of fiber, satellite, and microwave links. For instance, in cISP~\cite{cisp}, the authors demonstrated the competitive performance of a microwave link between Washington, DC, and New York City. Industry actors like Starlink~\cite{starlink}, Taara~\cite{taara}, Kuiper~\cite{kuiper}, or TeleSat~\cite{telesat} have developed new commercial Internet Service Providers (ISP) serving customers through wireless links. On top of a more direct path, these alternatives also benefit from shorter propagation delays compared to fibers.

While these hybrid deployments outperform fiber-only backbones in optimal scenarios, they are subject to high environmental-induced packet loss: weather events~\cite{cisp, wirelessFinancial}, physical objects blocking the line of sight path~\cite{LOSBlocking}, and human-induced interferences~\cite{interference} impact signal-to-noise ratio and result in transient packet loss surges~\cite{run-walk-hotnets, radwan_sigcomm, optical_failures_imc}.

\begin{figure}
    \centering
    \includegraphics[width=\linewidth]{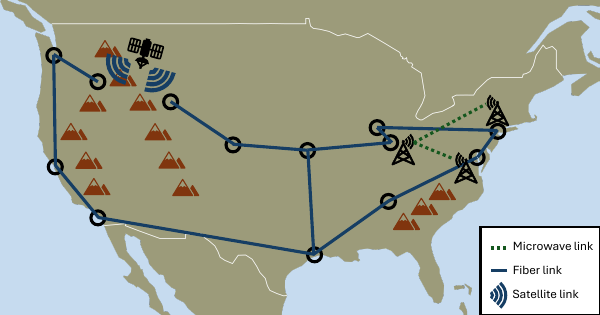}
    \caption{An illustration of a hybrid wireless-fiber backbone. The topology is inspired from Topology Zoo~\cite{topoZoo}.}
    \Description[Illustration of a hybrid wireless-fiber backbone]{A diagram of the United States with a backbone network composed of fiber, microwave, and satellite links.}
    \label{fig:illustration}
\end{figure}

Conventionally, end-hosts implement transport protocols to recover from network packet drops.
The choice of the transport protocol is, therefore, agnostic to the types of links used in the backbone networks.
Most transport protocols in today's backbones~\cite{covidTransportStudy} treat loss as a signal for congestion. Therefore, environmental-induced packet loss translates into reduced sender rates and link under-utilization.

To address the above challenge, prior work proposed using Network Coding (NC)~\cite{ncSdn, rlnc, slidingWindowCoding, ncLatencySurvey}. NC offers strong resiliency to random packet loss and provides low-latency guarantees.
However, prior NC approaches assume access to the end-host network stack~\cite{TCPNC, ctcp, acrlnc-sp, AdaptiveCausalNC}.

In this paper, we argue that this assumption does not always hold, particularly in hybrid wireless-fiber networks. We posit that sans access to end hosts, encoding and decoding operations must be performed somewhere else on the path. Consequently, we propose a novel approach to tame the impact of environmental-induced packet loss on reliable congestion control protocols without requiring changes to the end hosts. 

Our system, called \name, performs in-network NC while only assuming control over arbitrarily small subsets of the network switches. \name relies on capabilities of programmable network switches, such as Intel Tofino~\cite{tofino}, to implement a simple yet effective erasure coding mechanism inside network switches.

To perform NC, \name uses a systematic block coding approach that we detail in section~\ref{sec:model}. Our theoretical analysis quantifies \name's goodput gains by deriving the probability of retransmitting a packet with and without \name. 
We quantify how the choice of coding parameters impacts the rate of retransmissions and the network goodput (\S\ref{sec:modelAnalysis}). 

We evaluate \name in simulations on real-world backbone topologies and demonstrate that \name eliminates packet retransmission due to environmental packet loss, therefore cutting E2E latency by up to $18\%$.

\section{Background and Related Work}\label{sec:motivation}

{\textbf{Satellite and millimeter wave links.}}
Several ongoing industrial efforts are underway to provide connectivity in areas where fiber deployments are challenging. For example, Facebook connectivity proposed Magma~\cite{Magma} as an effort to improve connectivity of remote rural communities, and TerraGraph~\cite{terragraph} for last-mile connectivity in dense urban areas through millimeter wave links. Similarly, Taara~\cite{taara} proposes beam-of-light communications for high-speed, high-capacity connections in areas where fiber deployments are not economically viable. Satellite-based commercial ISPs have existed for some time (see for example HughesNet~\cite{hughesNet} and ViaSat~\cite{viaSat}) but suffered from high latency due to the altitude of their orbit. Recent low-earth orbit constellations like Starlink~\cite{starlink}, Kuiper~\cite{kuiper}, or Telesat~\cite{telesat} report E2E latencies in the order of tens of ms.

{\textbf{Environmental-induced packet loss.}} Environmental-induced packet loss in backbone networks is fundamentally linked to the physical vulnerabilities of optical and hybrid systems. Ghobadi et al. identified that backbone networks are highly susceptible to environmental disruptions, such as temperature fluctuations, which degrade signal quality over time, and physical incidents like fiber cuts or conduit damage during storms, leading to outages that cascade across channels sharing the affected segment~\cite{optical_failures_imc}. Singh et al. observe that variations in signal-to-noise ratio, even under controlled conditions, can be amplified by external factors like weather events and infrastructure aging, causing significant signal degradation~\cite{run-walk-hotnets, radwan_sigcomm}. Recently, Myers et al. showed that transient phenomena like rain, fog, atmospheric interference, and unplanned maintenance further exacerbate the instability of microwave-based links~\cite{wirelessFinancial}. These environmental impacts often trigger packet loss spikes by crossing critical signal-to-noise thresholds, making recovery challenging and highlighting the inherent fragility of hybrid wireless-fiber networks to environmental-induced packet loss events.

{\textbf{Congestion control.}} Traditional E2E congestion control algorithms, like TCP, rely on packet drops occurring at full network buffers as a signal for congestion.
This approach is based on the assumption that only congestion is responsible for dropped packets, which is reasonable for wired links with low erasure probability.
For paths containing a wireless link, however, since TCP is unable to discriminate between packet loss due to channel erasures or due to congestion, the channel capacity will often be significantly underestimated. 
This shortfall of TCP in high-loss systems has been well-studied by the community~\cite{linkLayerRetransmission, tcpWireless, tcpWirelessBwDx, tcpWirelessReport, tcpFEC}.

{\textbf{Network coding.}} Among possible approaches, NC has shown to be an efficient solution to the above problem. First introduced as a capacity-achieving technique~\cite{networkInformationFlow}, NC has been extended and improved in various ways~\cite{rlnc, slidingWindowCoding, AdaptiveCausalNC}, providing high-bandwidth and low latency guarantees. In~\cite{TCPNC}, authors have shown how to augment the TCP/IP stack with NC by adding a coding layer. This proposal also implemented re-encoding at intermediate nodes, providing flexibility to variations in the erasure probability on different links. In~\cite{ncSdn}, authors demonstrated the feasibility of NC approaches in software-defined networks and showed again the benefits of re-encoding at intermediate nodes.

Overall, these proposals have demonstrated the potential of NC. However, they all rely on access to and control of end-hosts. While encoding and decoding packets in-network has a latency cost compared to other E2E NC approaches, we argue in this paper that NC should also be used in cases where the network provider does not have access to end-hosts and can only modify subsets of the backbone network.

\section{System Design and Coding Approach}\label{sec:modelAnalysis}

In this section, we first present \name's design and assumptions~(\S\ref{sec:model}), before deriving the probability of retransmitting a packet in \name and its impact on aggregate packet arrival rate~(\S\ref{sec:goodput_analysis}).

\subsection{\name System Design}\label{sec:model}

\begin{figure}[t]
    \centering
    \includegraphics[width=0.9\linewidth]{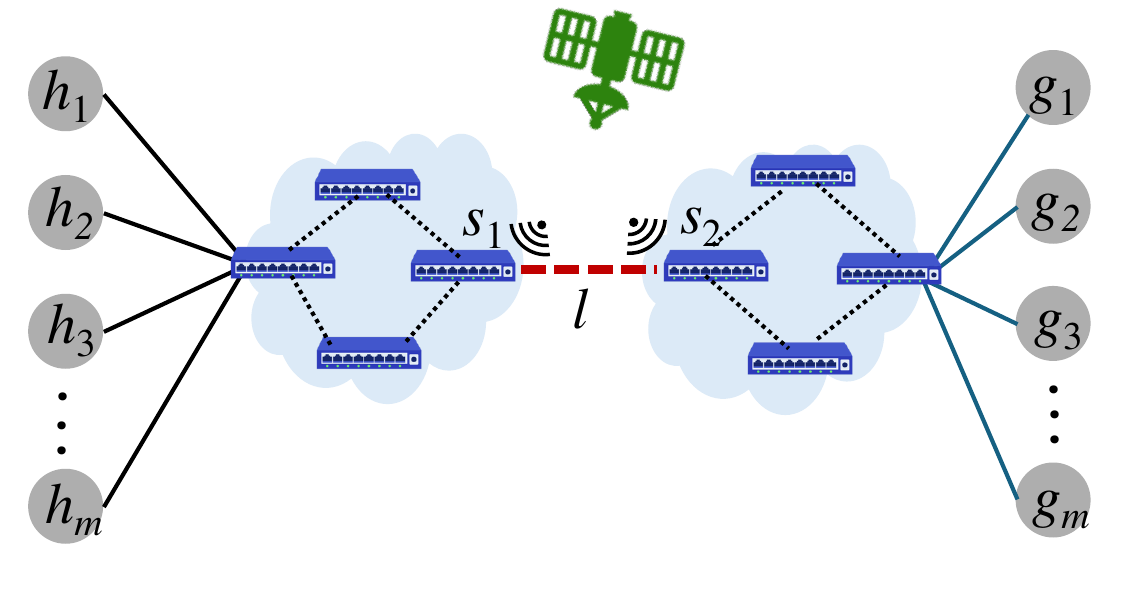}
    % \vspace{-.4cm}
    \caption{Illustration of \name's system model.}
    \label{fig:linc_model}
\end{figure}

Figure~\ref{fig:linc_model} illustrates \name's system model. We consider a set of hosts sending data through a network to a set of receivers. Among them, senders $h_1$, $h_2$, \dots, $h_m$ share a lossy link, $l$, on their path to their corresponding receivers $g_1$, $g_2$, ..., $g_m$. For any sender $h_i$, we define $\eta_i$ to be the number of non-lossy links on the ($h_i$, $g_i$) path. We model link $l$ as a packet erasure channel~\cite{ip_over_wireless_book}, with an associated environmental loss probability $\epsilon$. We assume that losses are not correlated.

Each source $h_i$ generates packets according to a Poisson process with an average of $\lambda_i$ packets per second. Let $s_1$ and $s_2$ be switches connected by link $l$, and consider the case where traffic flows from $s_1$ to $s_2$. In practice, acknowledgment packets traverse the other way; our analysis extends naturally to reverse traffic as well. % We assume that both switches are output-queued and have a large buffer space. 

We take the viewpoint of a network operator, who only has control over some parts of the network, including $s_1$ and $s_2$, but not over users' devices or over applications running on these devices. As these applications set transport protocols, we assume that traffic uses some version of TCP~\cite{covidTransportStudy} that we do not have control on. Note that despite this assumption, our method is transparent to end-hosts, and therefore compatible with any other E2E coding protocols. 

{\textbf{In-network encoding.}} \name uses a systematic block code to encode packets at $s_1$. More specifically, for every block $B = (P_1, P_2, ..., P_k)$ of $k$ consecutive incoming packets to $s_1$ with $s_2$ as the next hop, $s_1$ first transmits all packets of $B$ unmodified. Then, $s_1$ generates $n-k$ additional packets destined to $s_2$, thereby sending a total of $n$ packets to $s_2$. The process of creating the coded packets in $s_1$ consists in calculating a linear combination of packets in block $B$, where the coefficients are chosen such that if any $k$ out of the $n$ packets are successfully transmitted, the resulting matrix will be invertible (\textit{i.e.} a maximum distance separable code~\cite{systematicCodes}).
\name's packet header includes $\log_2(n)$ bits to identify the index of the packet in the block. This allows switch $s_2$ to determine whether a packet has been coded, and if so which linear combination should be used to decode it. 
% \benoit{I could not find a result on whether there is a general bound on k for an MDS code to exist. Should we specify a maximum value for $n-k$? Is there a bound in general?}

{\textbf{In-network decoding.}} Because of the environmental losses on link $l$, switch $s_2$ receives a subset of the $n$ packets sent by $s_1$. Switch $s_2$ forwards successfully received uncoded packets to their next hops without waiting for all the packets in the block to be transmitted.
For every block, $s_2$ counts the number of successfully received packets. If at least $k$ packets are received, $s_2$ decodes the block and starts sending the lost packets that it recovered. If less than $k$ packets are received, $s_2$ moves on to the next block, so that the transport layer reliability mechanism can detect losses and trigger retransmissions.

\subsection{\name Theoretical Analysis}
\label{sec:goodput_analysis}

Consider the backbone network shown in Figure~\ref{fig:linc_model}, where link $l$ drops packets with probability $\epsilon$. As discussed in the previous section, \name recovers lost packets using its in-network encoding/decoding techniques. As a result, a \name backbone network has a higher goodput compared to a backbone without \name. In this section, we derive mathematical formulations to quantify \name's goodput gains.

% In the model described in section~\ref{sec:model}, senders generate packets according to a Poisson process. In switches upstream up to $s_1$, assuming exponentially distributed transmission, propagation, and processing delays, the arrival process of packets corresponds to a combination of Poisson processes, therefore also a Poisson process. The output process of the decoder at $s_2$ is not Poisson since several packets are decoded and sent when $k$ packets are received. This traffic, however, is mixed in the rest of the network with packets from a large number of other sources, and can therefore be approximated to a Poisson process using the Kleinrock assumption~\cite{kleinrockThesis}.
\subsubsection{Retransmission Rate}
In the block coding approach we described in section~\ref{sec:model}, a block is successfully decoded if at least $k$ out of $n$ packets in the block are successfully transmitted.
% Given our assumption of non-correlated losses, the probability of a block being unrecoverable $\pi$ is therefore given by:
% \begin{align}
%     \pi &= \mathbbm{P}(L> n-k)\nonumber\\
%     & = \sum_{i=n-k+1}^{n} \mathbbm{P}(L = i)\nonumber\\
%     & = 1 - \sum_{i=0}^{n-k} \mathbbm{P}(L = i)\nonumber\\
%     & = 1 - \sum_{i=0}^{n-k} \binom{n}{i} \epsilon^i (1-\epsilon)^{(n-i)}
% \end{align}
% with $L$ the number of losses in a block.
Given that our block code is systematic, when a block is not recoverable all lost uncoded packets must be retransmitted. Note that uncoded packets correspond to the original data packets and need to be retransmitted, whereas we do not need to retransmit coded packets that are lost. Let $Q$ be the random variable corresponding to the number of losses in a block, and $C$ be the random variable corresponding to the number of lost uncoded packets.

To quantify \name's goodput gains, we start by deriving the expected packet retransmission rate $R_{\name}$, by formulating the expected value of the number of uncoded packets that have been lost (i.e., $\mathbbm{E}[Q, C]$), as follows:

\begin{align}
    R_{\name} &= \frac{\mathbbm{E}[Q, C]}{k} \nonumber\\
    &= \frac{1}{k}\sum_{q=n-k+1}^n \sum_{c=q-(n-k)}^{\min(k, q)} c \cdot \mathbbm{P}(Q=q) \mathbbm{P}(C=c \ | \ Q=q) \nonumber\\
    & = \frac{1}{k}\sum_{q=n-k+1}^n \sum_{c=q-(n-k)}^{\min(k, q)} c \cdot \binom{n}{q} \epsilon^q (1-\epsilon)^{(n-q)}\frac{\binom{k}{c}\binom{n-k}{q-c}}{\binom{n}{q}}\nonumber\\
    & = \frac{1}{k}\sum_{q=n-k+1}^n \sum_{c=q-(n-k)}^{\min(k, q)} c \cdot \binom{k}{c}\binom{n-k}{q-c} \epsilon^q (1-\epsilon)^{(n-q)}\nonumber\\
    & = \frac{1}{k}\sum_{q=n-k+1}^n \epsilon^q (1-\epsilon)^{(n-q)} \sum_{c=q-(n-k)}^{\min(k, q)} c \cdot \binom{k}{c}\binom{n-k}{q-c}\label{eq:epsilonLINC}
\end{align}
In the above equation, $R_{\name}$ is computed by taking the expected value of the joint distribution $(Q, C)$. Using the law of total probability, we express this joint distribution as a function of $Q$ and the conditional distribution of $C$, conditioned on $Q$. Since packet losses are independent, $Q$ follows a binomial distribution. To determine the distribution of $C$, we draw the following analogy: in a block of $n$ packets, there are $k$ uncoded packets, and $n-k$ coded ones. We randomly select $q$ packets, and try to determine the number of uncoded packets among these $q$. As a result, $(C\ | \ Q)$ corresponds to a hypergeometric distribution. 
\subsubsection{Aggregate Packet Arrival Rate}
\label{sec:agg_packet_arrival_rate}
Next, we formulate the aggregate packet arrival rate over all links in the network. To do so, we first formulate the aggregate arrival rate of packets on link $l$ ($\lambda_{\name}$) by considering the following three categories of traffic on link $l$: ($i$) the aggregate raw traffic from all the senders $(\sum_{i=1}^m \lambda_i)$, ($ii$) added coded packets ($\frac{n}{k}$), and ($iii$) the number of retransmissions for lost packets that were not recovered. The probability of a packet being retransmitted $s$ times is given by $R_{\name}^s$, thus the fraction of retransmitted packets is given by the sum of $R_{\name}^s$ over the number of retransmissions. Putting it together, we have: 
\begin{align}
    \lambda_{\name} &= \left(\sum_{i=1}^m \lambda_i \right)\cdot \left(\frac{n}{k}\right) \cdot \left(\sum_{s=0}^{+\infty} R_{\name}^s\right)  = \frac{n \sum_{i=1}^m \lambda_i}{k(1-R_{\name})}
\end{align}

Second, we formulate the aggregate arrival rate of packets on all the other non-lossy links in the network ($\lambda_{\name}'$). As we are summing over all non-lossy link, the traffic from each sender contributes to the arrival rate of $\eta_i$ links, hence:
\begin{align}
    \lambda'_{\name} &= \left(\sum_{i=1}^m \eta_i \lambda_i \right) \cdot \left(\sum_{s=0}^{+\infty} R_{\name}^s\right) = \frac{\sum_{i=1}^m \eta_i\lambda_i}{1 - R_{\name}}
\end{align}

% Hence a goodput of:
% \begin{align*}
%     \frac{\sum_{i=1}^m \lambda_i}{\lambda_{\name}} = \frac{k-R_{\name}}{n}
% \end{align*}

The total aggregate packet arrival rate on a \name network is $\lambda_{\name} +\lambda'_{\name}$.

\subsubsection{Goodput Gains}
Finally, to derive \name's goodput gains, we first repeat the above calculations for the case without \name. Then, we derive the ratio of goodputs for systems with and without \name.

In a network without \name, the retransmission rate is given by $R_{noNC} = \epsilon$, yielding a total arrival rate of packets on the lossy link, as:
\begin{align}
    \lambda_{noNC}&=\left(\sum_{i=1}^m \lambda_i \right)\sum_{s=0}^{\infty}\epsilon^s = \frac{\sum_{i=1}^m \lambda_i }{1-\epsilon}
\end{align}

Similarly, on the other links:
\begin{align}
    \lambda'_{noNC}&=\left(\sum_{i=1}^m \eta_i\lambda_i \right)\sum_{s=0}^{\infty}\epsilon^s = \frac{\sum_{i=1}^m \eta_i\lambda_i}{1-\epsilon}
\end{align}
Goodput is defined as the ratio of useful packets (the raw traffic from all senders traveling over $h_i + 1$ links for sender~$i$) over the total number of packets transmitted. The goodput for a system using \name, $\mathcal{G}_{\name}$, (respectively $\mathcal{G}_{noNC}$ for a system not using \name) is given by:
\begin{align*}
    \mathcal{G}_{\name} = \frac{\sum_{i=1}^m \lambda_i (h_i + 1)}{\lambda_{\name} + \lambda_{\name}'} &&&     \mathcal{G}_{noNC} = \frac{\sum_{i=1}^m \lambda_i (h_i + 1)}{\lambda_{noNC} + \lambda_{noNC}'}
\end{align*}

Finally, we define $\Delta$, the ratio of goodputs for a system with \name over a system without \name:

\begin{align}
    \Delta &= \frac{\lambda_{noNC} + \lambda'_{noNC}}{\lambda_{\name} + \lambda'_{\name}} \nonumber
    = \frac{\frac{\sum_{i=1}^m \lambda_i }{1-\epsilon} + \frac{\sum_{i=1}^m \eta_i\lambda_i}{1-\epsilon}}{\frac{n \sum_{i=1}^m \lambda_i}{k(1-R_{\name})} + \frac{\sum_{i=1}^m \eta_i\lambda_i}{1 - R_{\name}}}\nonumber\\
    & = \frac{1-R_{\name}}{1-\epsilon}\frac{\sum_{i=0}^m \lambda_i \left(\eta_i + 1\right)}{\sum_{i=0}^m \lambda_i \left(\eta_i+\frac{n}{k}\right)}\label{eq:delta}
\end{align}

To determine the optimal choice of coding parameters ($k$, $n$) for \name, we propose an optimization formulation that maximizes $\Delta$ subject to the following constraints: ($i$) $n \geq k$ and ($ii$) $k > 0$. The optimization takes the network topology, end-hosts sending rates $\lambda_i$ and $\epsilon \in [0, 1]$ as input parameters.

\section{Evaluations}\label{sec:evals}

\begin{figure}[h]
     \centering
     \begin{subfigure}[b]{0.25\linewidth}
         \centering
         \includegraphics[width=\textwidth]{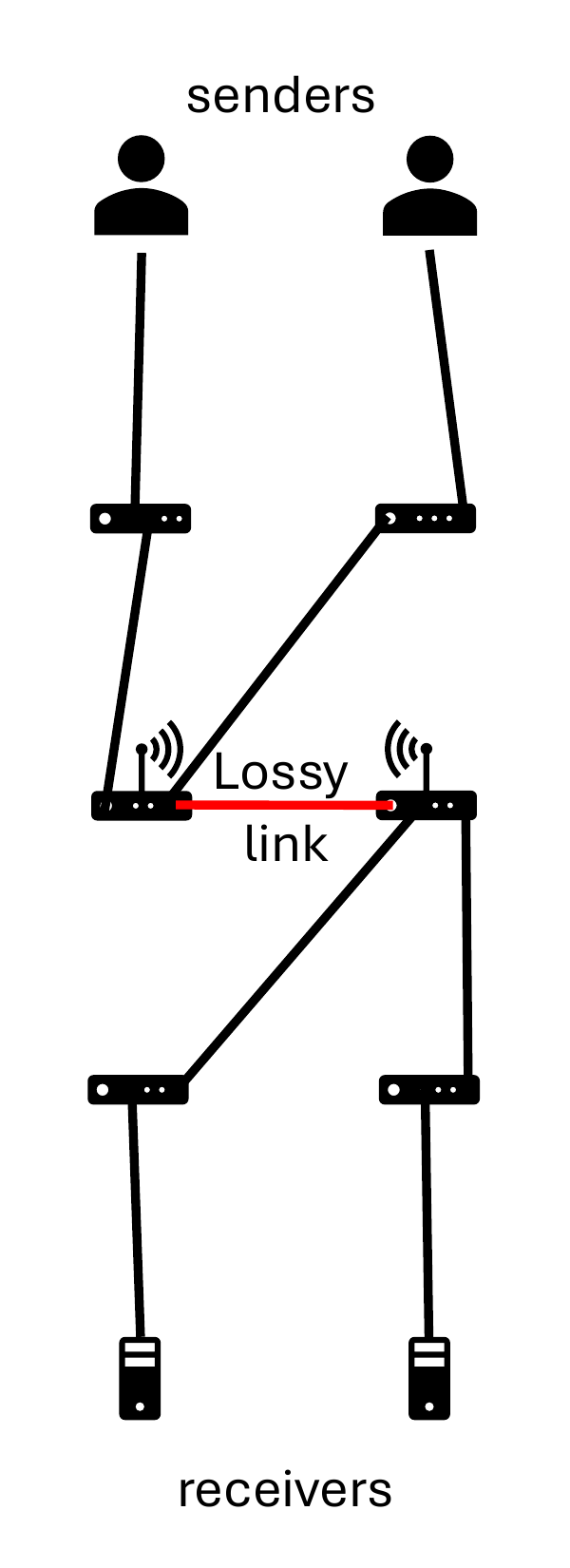}
         \caption{Scenario 1}
         \label{fig:scenario1}
     \end{subfigure}
     \hfill
     \begin{subfigure}[b]{0.60\linewidth}
         \centering
         \includegraphics[width=\textwidth]{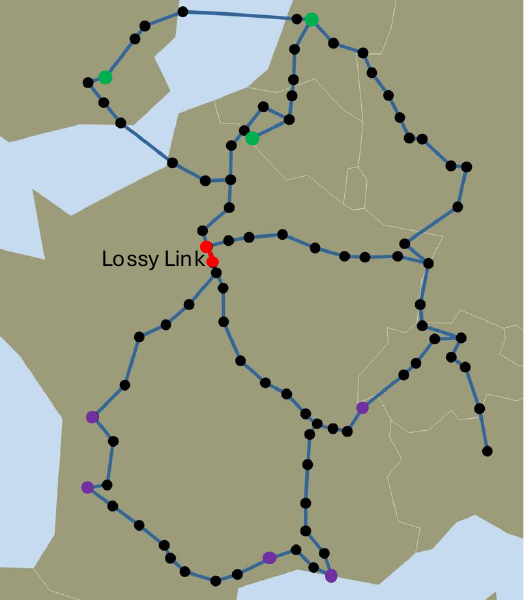}
         \caption{Scenario 2: VtlWavenet2011~\cite{topoZoo}.}
         \label{fig:scenario2}
     \end{subfigure}
        \caption{Topologies used in our simulations.}
        \label{fig:topologies}
\end{figure}
\begin{figure*}
     \centering
     \begin{subfigure}[b]{0.226\linewidth}
         \centering
         \includegraphics[width=\linewidth]{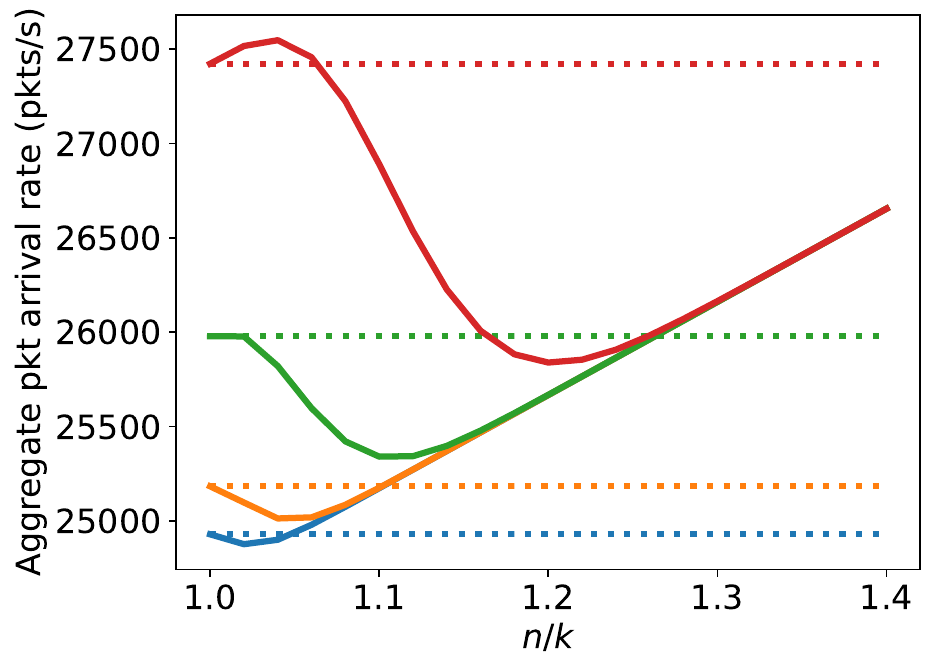}
         \caption{Scenario 1, Theoretical}
         \label{fig:epsilon-th-sc1}
     \end{subfigure}
     \hfill
     \begin{subfigure}[b]{0.216\linewidth}
         \centering
         \includegraphics[width=\linewidth]{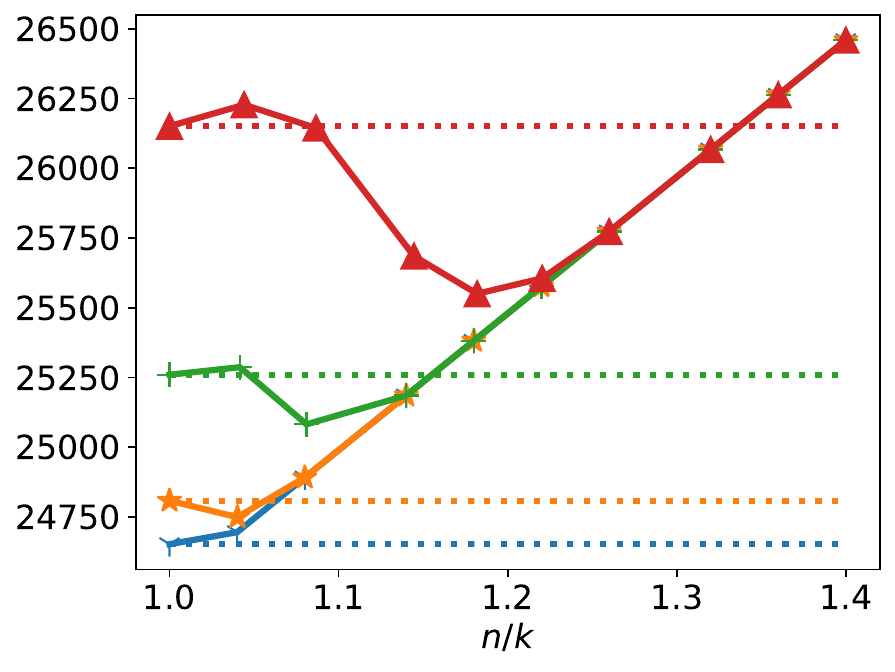}
         \caption{Scenario 1, Simulations}
         \label{fig:epsilon-emp-sc1}
     \end{subfigure}
     \hfill
     \begin{subfigure}[b]{0.221\linewidth}
         \centering
         \includegraphics[width=\linewidth]{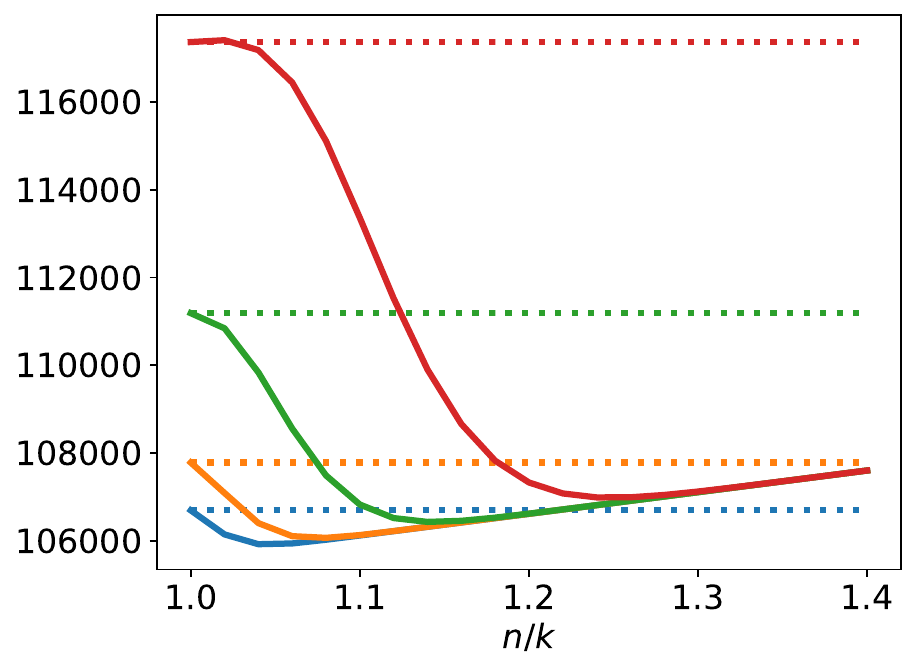}
         \caption{Scenario 2, Theoretical}
         \label{fig:epsilon-th-sc2}
     \end{subfigure}
     \hfill
     \begin{subfigure}[b]{0.286\linewidth}
         \centering
         \includegraphics[width=\linewidth]{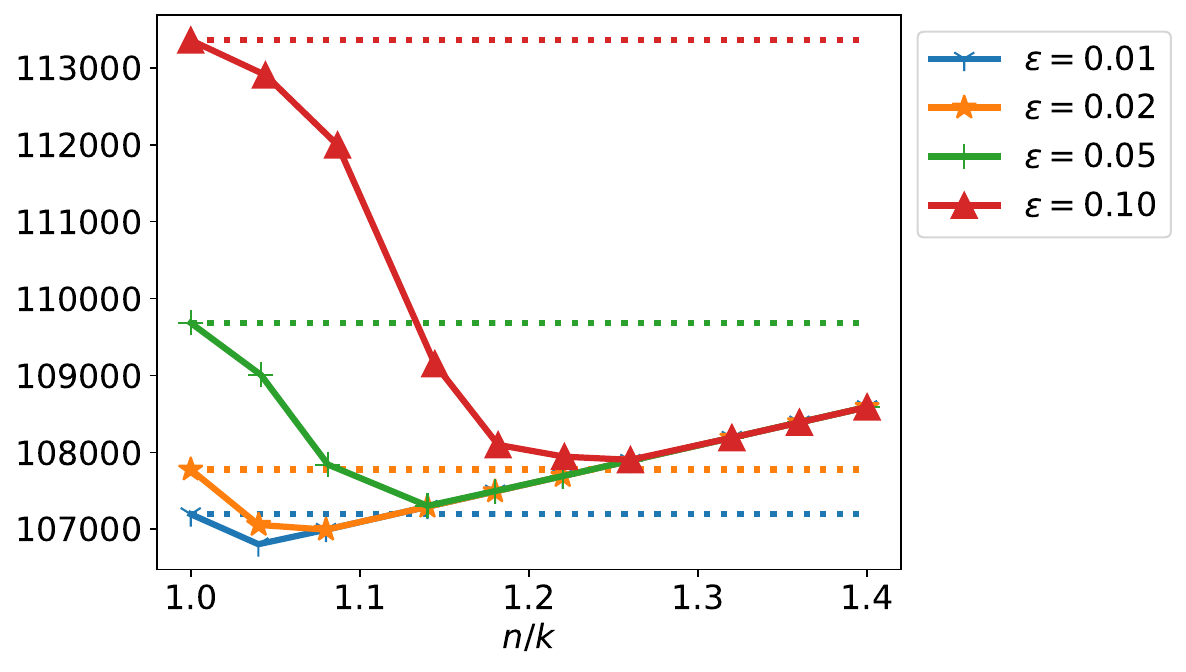}
         \caption{Scenario 2, Simulations}
         \label{fig:epsilon-emp-sc2}
     \end{subfigure}
        \caption{Aggragate packet rate for different erasure probabilities $\epsilon$ of the lossy link.}
        \label{fig:evalEpsilon}
\end{figure*}

\begin{figure*}
     \centering
     \begin{subfigure}[b]{0.227\linewidth}
         \centering
         \includegraphics[width=\linewidth]{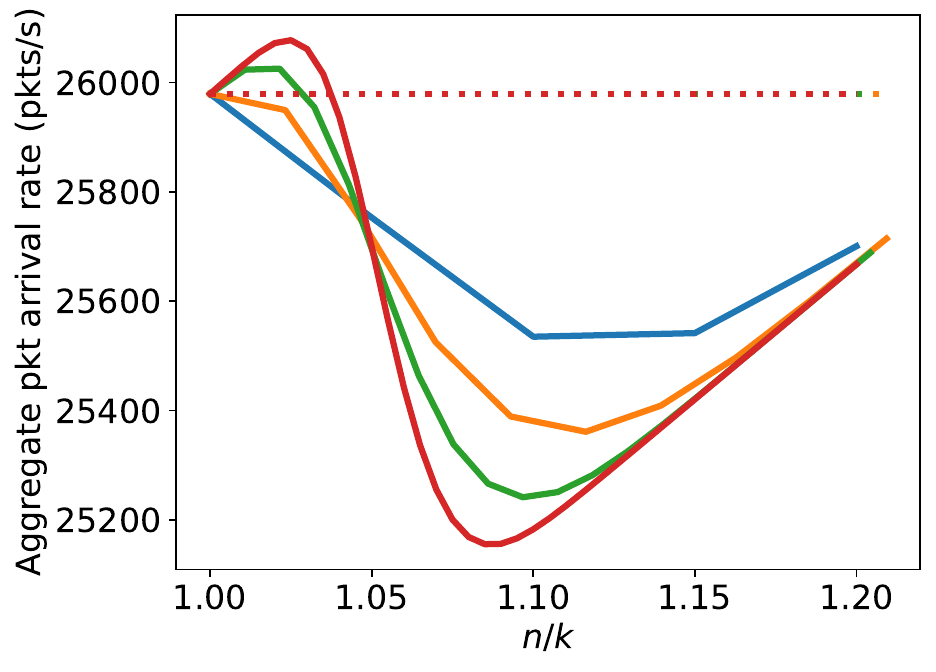}
         \caption{Scenario 1, Theoretical}
         \label{fig:k-th-sc1}
     \end{subfigure}
     \hfill
     \begin{subfigure}[b]{0.218\linewidth}
         \centering
         \includegraphics[width=\linewidth]{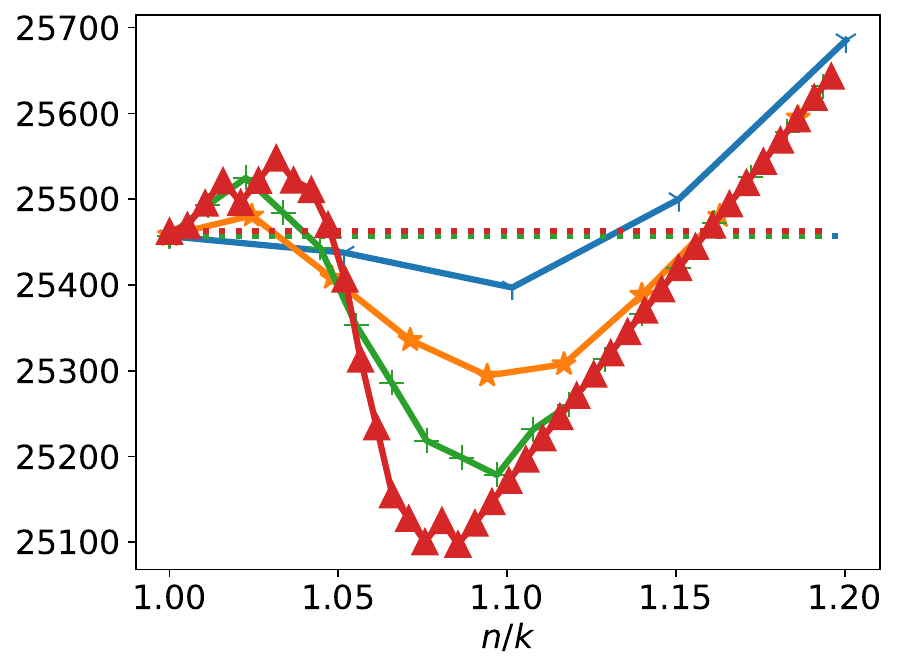}
         \caption{Scenario 1, Simulations}
         \label{fig:k-emp-sc1}
     \end{subfigure}
     \hfill
     \begin{subfigure}[b]{0.222\linewidth}
         \centering
         \includegraphics[width=\linewidth]{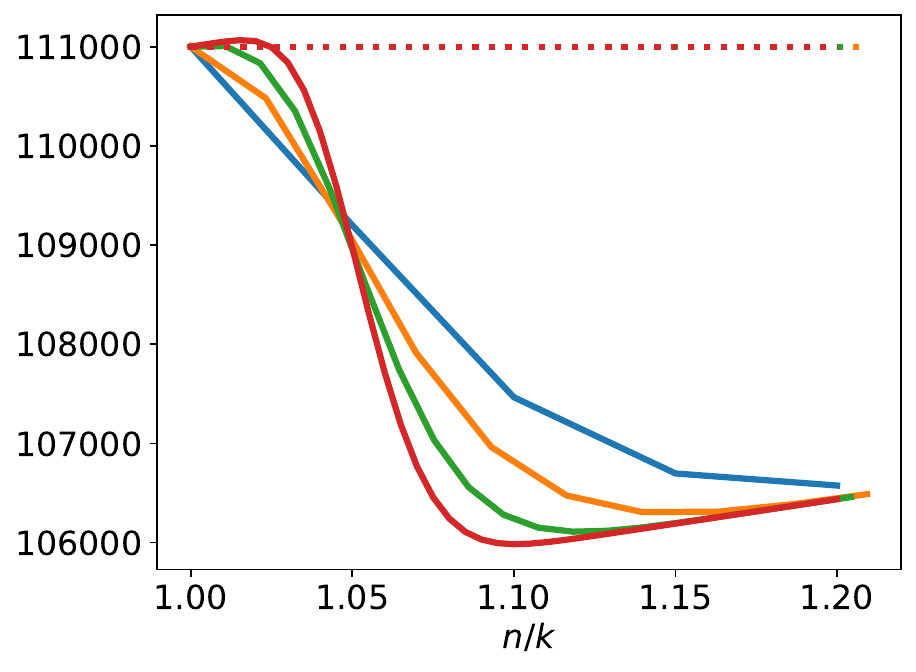}
         \caption{Scenario 2, Theoretical}
         \label{fig:k-th-sc2}
     \end{subfigure}
     \hfill
     \begin{subfigure}[b]{0.283\linewidth}
         \centering
         \includegraphics[width=\linewidth]{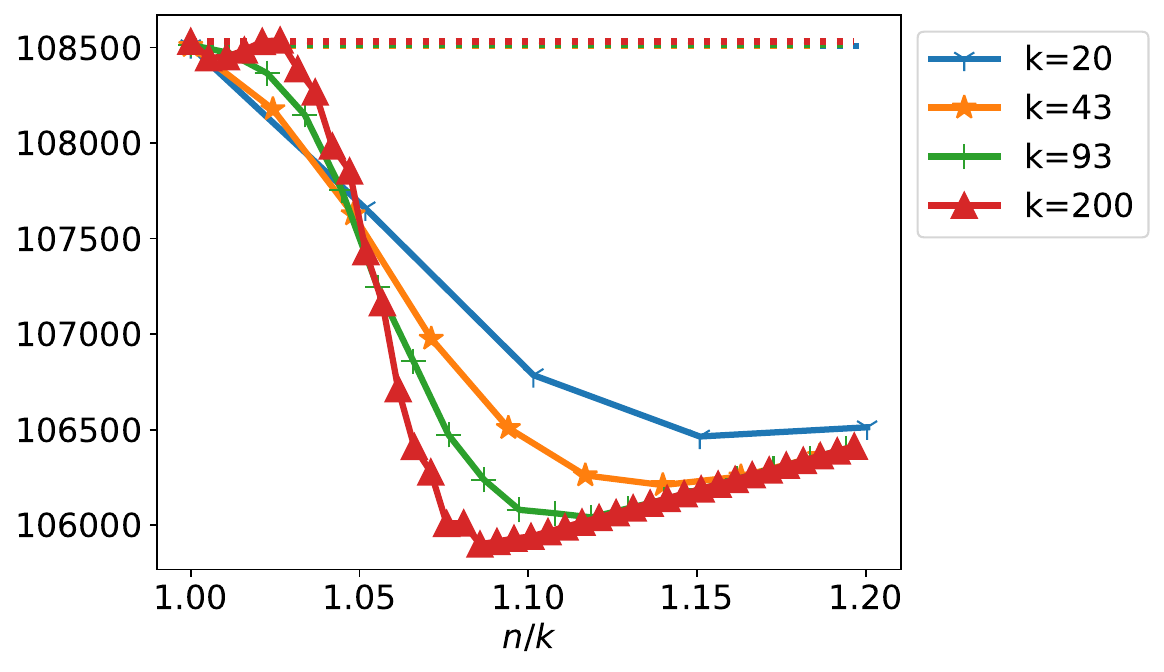}
         \caption{Scenario 2, Simulations}
         \label{fig:k-emp-sc2}
     \end{subfigure}
        \caption{Aggragate packet rate for different values of $k$.}
        \label{fig:evalk}
\end{figure*}

To evaluate \name, we implement a custom event-based simulator. This section describes our simulation methodology and compare its performance with the state-of-the-art.
\subsection{Methodology}\label{sec:evalsMethodology}

{\textbf{Simulator.}} Our simulation generates packets on senders following Poisson processes. The sending rates are randomly selected across senders and are normalized to an average network load of 50\%. To simulate environmental packet loss, we select a single link to randomly drop packets. Each link is modeled by a shared resource processing packets. When a packet is lost, we trigger a retransmission after three duplicate ACKs are received by the sender. We measure several parameters including the number of lost packets that were recovered using \name, the number of retransmitted packets, and the arrival rate of packets at each link. We define the E2E packet delay as the difference between the time a packet was received successfully on a receiver, and the time the same packet was created at the corresponding sender.

\textbf{{Topologies.}} We consider two topologies, shown in Figure~\ref{fig:topologies}. In the first scenario (Figure~\ref{fig:scenario1}), two senders are sending traffic over a 5-hop network, each to a different receiver. The middle link is lossy, with latency 1~ms and drop probability 5\%, unless otherwise stated. The latency for each non-lossy link is set to 100~ms. The number of non-lossy links for either sender is $4$, as shown in Figure~\ref{fig:scenario2}. The second scenario (Figure~\ref{fig:scenario2}) is the VtlWavenet2011 topology taken from the Topology Zoo dataset~\cite{topoZoo}. We set a central link near Paris to be the lossy link and choose the following (sender, receiver) pairs: (Marseille, London), (Bordeaux, Amsterdam), (Geneva, London), (Sete, Antoing), and (Blanzay, Strasbourg). This results in around $20$ non-lossy links for each (sender, receiver) pair. The latency for each link is computed from the geographical distance between nodes.
In both cases, all non-lossy links are assumed to have a rate of 100~Gbit/s. The lossy link is shown in red in both figures, and we highlight senders in purple and receivers in green in Figure~\ref{fig:scenario2}. For both scenarios, we set $k=50$ unless otherwise noted.

\begin{figure*}
     \centering
     \begin{subfigure}[b]{0.228\linewidth}
         \centering
         \includegraphics[width=\linewidth]{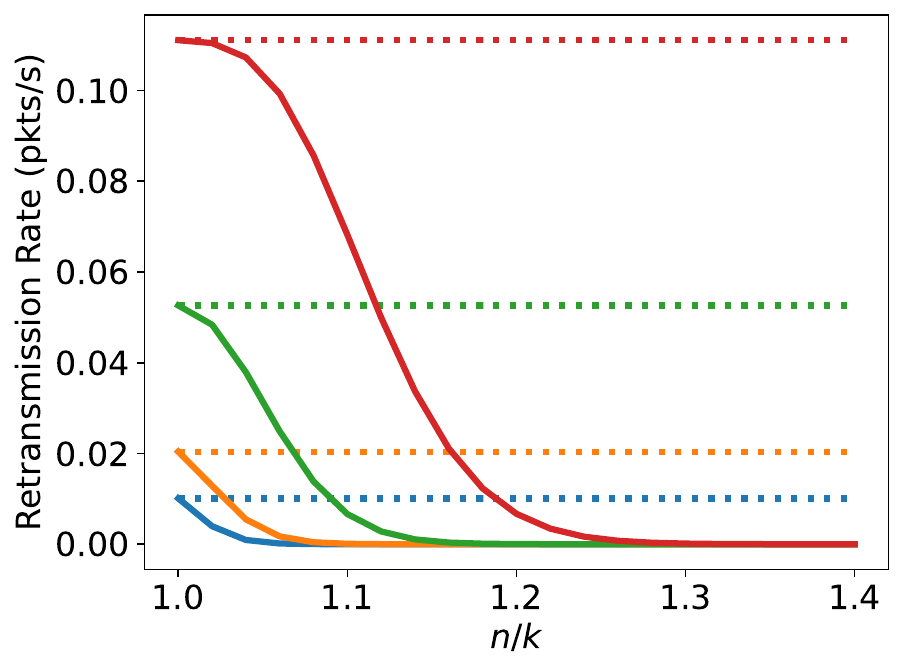}
         \caption{Scenario 1, Theoretical}
         \label{fig:epsilon-retrans-th-sc1}
     \end{subfigure}
     \hfill
     \begin{subfigure}[b]{0.218\linewidth}
         \centering
         \includegraphics[width=\linewidth]{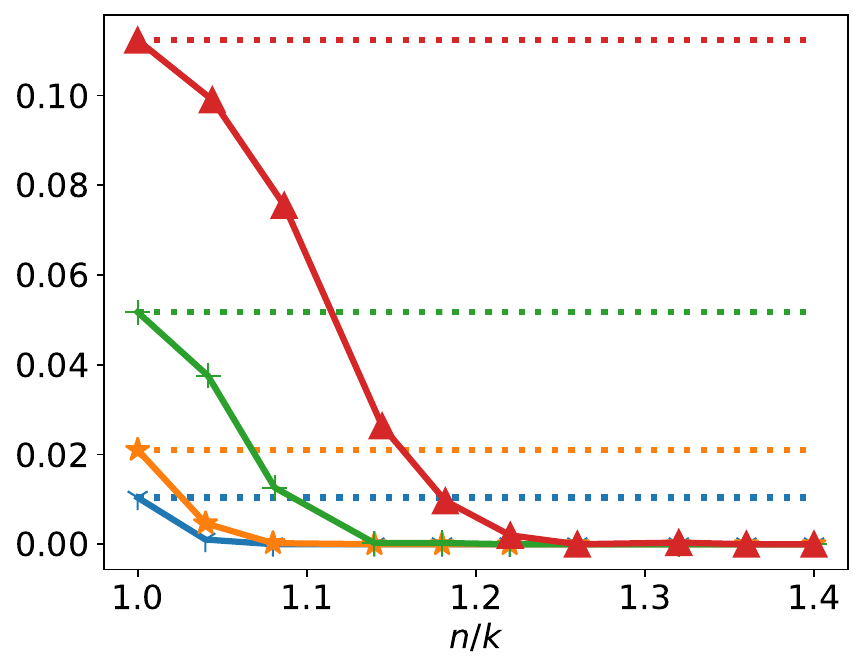}
         \caption{Scenario 1, Simulations}
         \label{fig:epsilon-retrans-emp-sc1}
     \end{subfigure}
     \hfill
     \begin{subfigure}[b]{0.218\linewidth}
         \centering
         \includegraphics[width=\linewidth]{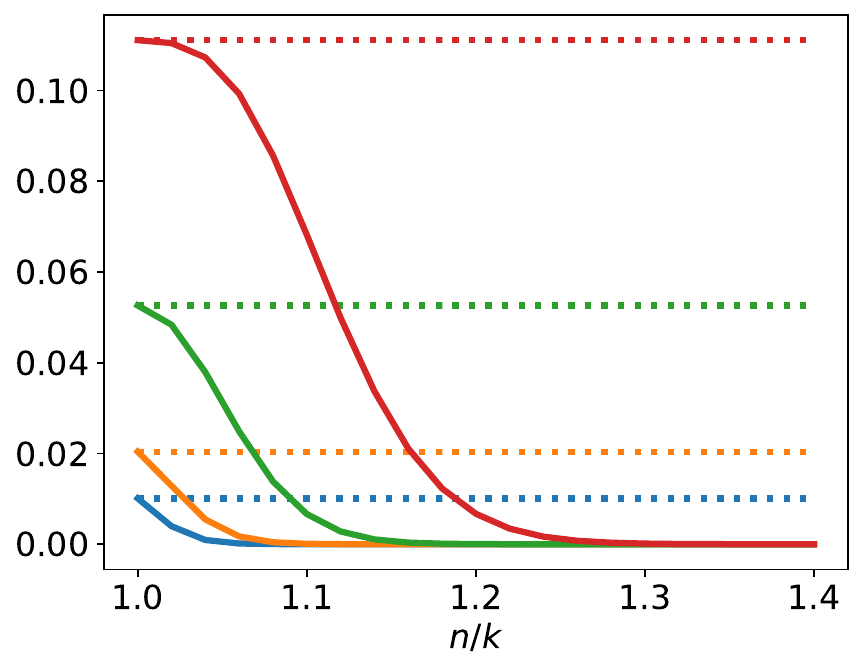}
         \caption{Scenario 2, Theoretical}
         \label{fig:epsilon-retrans-th-sc2}
     \end{subfigure}
     \hfill
     \begin{subfigure}[b]{0.286\linewidth}
         \centering
         \includegraphics[width=\linewidth]{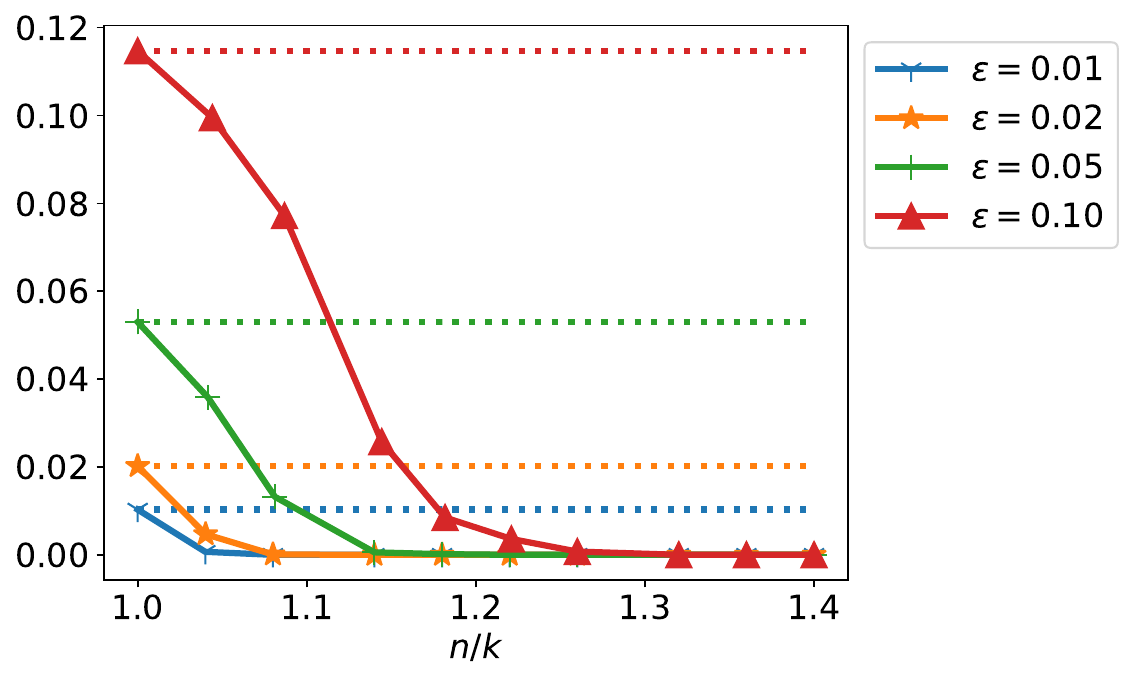}
         \caption{Scenario 2, Simulations}
         \label{fig:epsilon-retrans-emp-sc2}
     \end{subfigure}
        \caption{Rate of retransmissions for different erasure probabilities $\epsilon$ of the lossy link}
        \label{fig:evalEpsilonR}
\end{figure*}

\begin{figure*}
     \centering
     \begin{subfigure}[b]{0.229\linewidth}
         \centering
         \includegraphics[width=\linewidth]{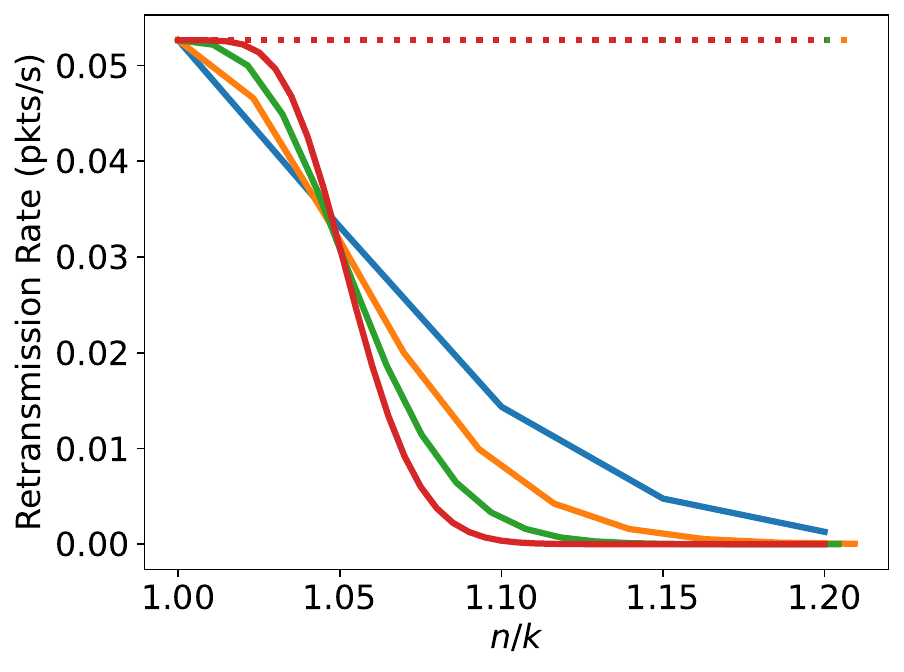}
         \caption{Scenario 1, Theoretical}
         \label{fig:k-retrans-th-sc1}
     \end{subfigure}
     \hfill
     \begin{subfigure}[b]{0.220\linewidth}
         \centering
         \includegraphics[width=\linewidth]{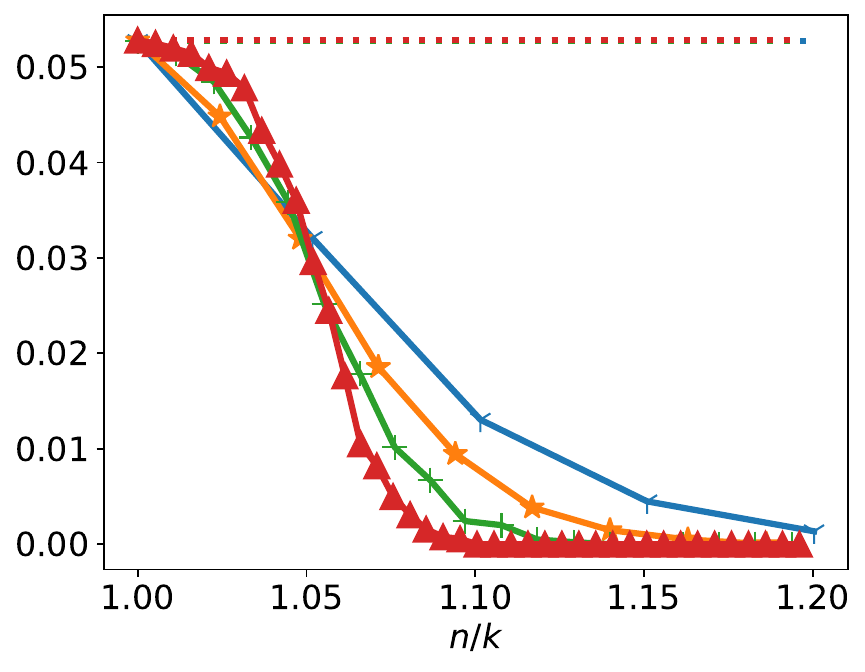}
         \caption{Scenario 1, Simulations}
         \label{fig:k-retrans-emp-sc1}
     \end{subfigure}
     \hfill
     \begin{subfigure}[b]{0.219\linewidth}
         \centering
         \includegraphics[width=\linewidth]{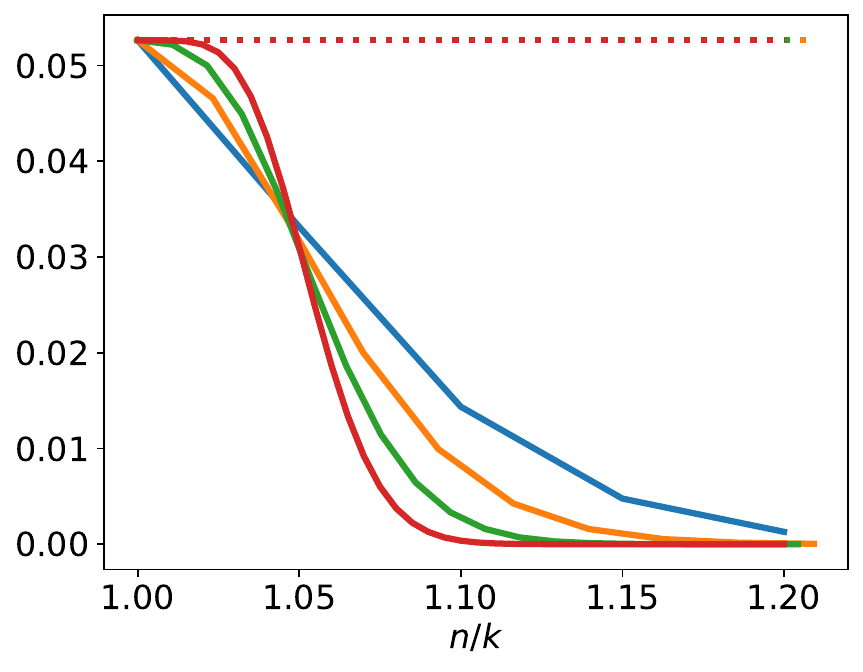}
         \caption{Scenario 2, Theoretical}
         \label{fig:k-retrans-th-sc2}
     \end{subfigure}
     \hfill
     \begin{subfigure}[b]{0.282\linewidth}
         \centering
         \includegraphics[width=\linewidth]{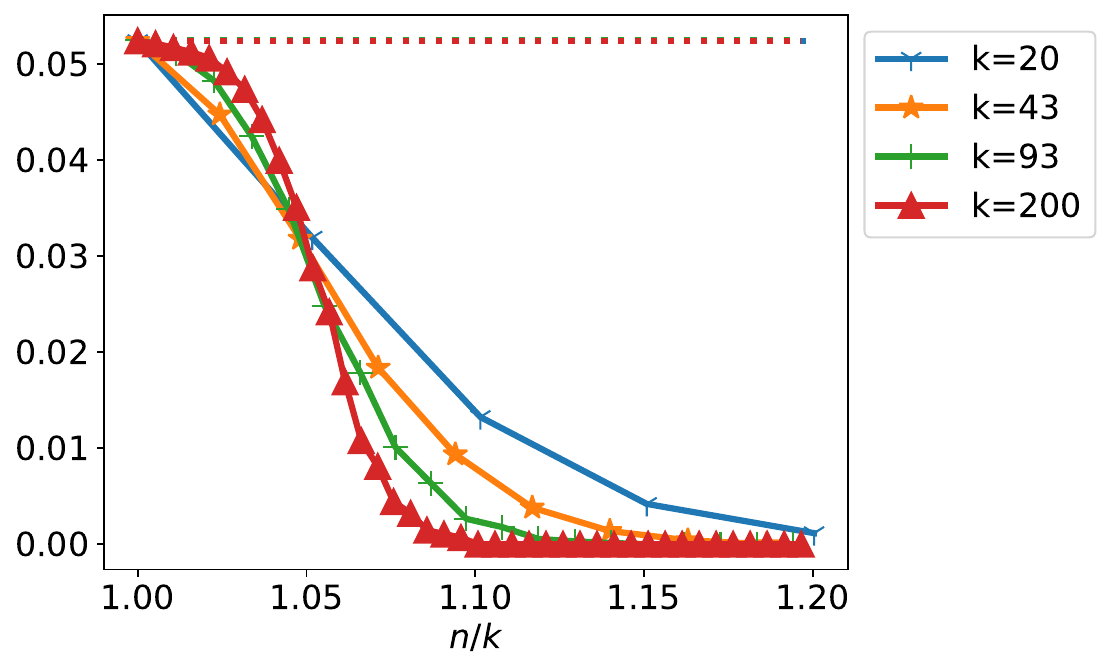}
         \caption{Scenario 2, Simulations}
         \label{fig:k-retrans-emp-sc2}
     \end{subfigure}
        \caption{Rate of retransmissions for different values of $k$.}
        \label{fig:evalkR}
\end{figure*}
\subsection{Aggregate Packet Arrival Rate}\label{sec:evalsAggregate}

In TCP, retransmissions happen E2E. Therefore, when a packet needs to be retransmitted, it utilizes bandwidth on every link on its path. Figures~\ref{fig:evalEpsilon} and~\ref{fig:evalk} compare the aggregate packet arrival rate of \name under different scenarios. 
In Figures~\ref{fig:epsilon-th-sc1} and~\ref{fig:epsilon-emp-sc1} (respectively Figures~\ref{fig:epsilon-th-sc2} and~\ref{fig:epsilon-emp-sc2}), we show that empirical results from our simulations closely align with our theoretical formulation (\S\ref{sec:agg_packet_arrival_rate}). The dotted lines represent the case without \name.

As expected, the minimum value for the aggregate packet rate is a function of the packet loss rate $\epsilon$. As $\epsilon$ increases, the amount of redundant packets needed to correct lossy transmissions increases. For instance, for $\epsilon=0.1$, Figure~\ref{fig:epsilon-emp-sc1} shows three regimes: ($i$) $1 \leq \frac{n}{k} < 1.04$ where the coding rate is too low to recover a meaningful number of packets, but extra coded packets contribute to the aggregate rate, ($ii$) $1.04 \leq \frac{n}{k} < 1.18$ where the retransmission rate decreases rapidly to 0, compensating for the added coded packets, and ($iii$) $1.18 \leq \frac{n}{k} < 1.4$ where no retransmissions remain, and increasing the coding rate only loads up the lossy link.
A similar behavior can be observed in both scenarios and across our experiments. 

The same behavior can be observed in both scenarios.

Figure~\ref{fig:evalk} shows similar results. Given the small number of hops in scenario 1, the relative impact of adding packets on the lossy link is more important than for scenario 2. This can be seen in the first and last regimes where the aggregate packet rate increases with the coding rate. Increasing $k$ has two effects: first, since $n$ and $k$ are integers, the granularity of choice for coding rates is finer when $k$ is large. If $k$ is too small, then the optimal coding rate might not be achievable. Second, our results show that even at a coding rate that can be achieved for all values of $k$ presented here, larger values of $k$ further reduce the aggregate packet rates.
As mentioned, our simulations show the same trend as predicted by our analytical model: in practical deployments, for a given $\epsilon$, the optimal coding rate can be computed using equation~\ref{eq:delta}.

\subsection{Retransmission Rate}
\label{sec:evalsRetransmission}

To evaluate the impact of reduction in retransmission probability in \name, Figures~\ref{fig:evalEpsilonR} and~\ref{fig:evalkR} plot the effects of $\epsilon$ and $k$ on retransmission probabilities.
In Figures~\ref{fig:epsilon-retrans-th-sc1} and~\ref{fig:epsilon-retrans-th-sc2}, we use equation~\ref{eq:epsilonLINC} to compute the expected rate of retransmissions and verify in Figures~\ref{fig:epsilon-retrans-emp-sc1} and~\ref{fig:epsilon-retrans-emp-sc2} that our simulations match the derivation.
Importantly, \name eliminates all unnecessary retransmissions by driving the retransmission rate to zero over the entire path, at the cost of a small added overhead on the single lossy link, for all values of $\epsilon$. 
In real-world implementations, service providers should use our theoretical model to determine the optimal coding rate in their setup.

\begin{figure}
     \centering
     \begin{subfigure}[b]{0.433\linewidth}
         \centering
         \includegraphics[width=\linewidth]{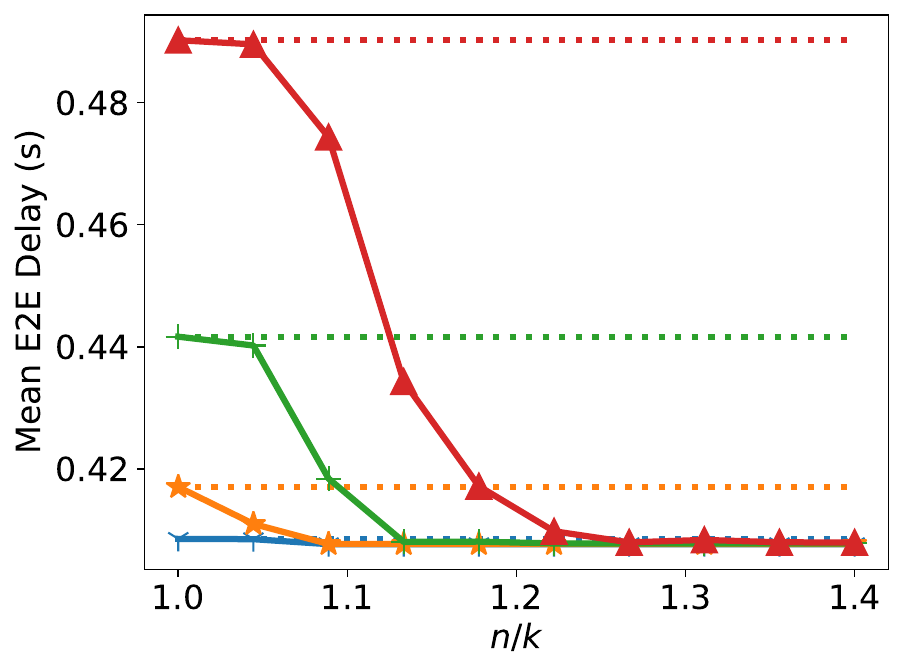}
         \caption{Scenario 1, Mean}
         \label{fig:mean-delay-sc1}
     \end{subfigure}
     \hfill
     \begin{subfigure}[b]{0.547\linewidth}
         \centering
         \includegraphics[width=\linewidth]{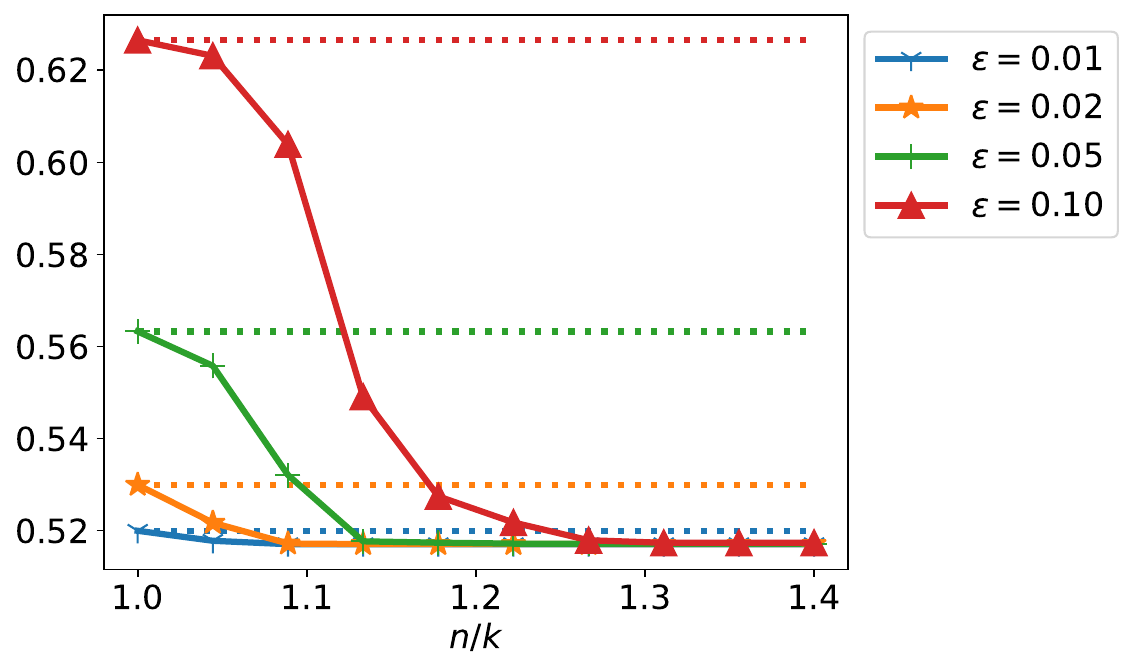}
         \caption{Scenario 2, Mean}
         \label{fig:mean-delay-sc2}
     \end{subfigure}
        \caption{E2E packet delivery delay.}
        \label{fig:evalDelay}
        \vspace{10pt}
\end{figure}
Figure~\ref{fig:evalkR} studies the impact of the value of $k$ on the rate of retransmissions. We observe the same trend as in Figure~\ref{fig:evalk}: the higher $k$, the closer the retransmission rate is to a step function. Again, we observe the same trends in simulation as predicted by our analysis.

\subsection{E2E Delay}\label{sec:evalsDelay}

Figure~\ref{fig:evalDelay} shows the mean E2E delay experienced by all packets in our simulations. \name reduces the delay by up to $18\%$, by reducing the amount of retransmissions required. For both scenarios, the average delay is primarily related to the probability of a packet being retransmitted, since a retransmission yields a delay of at least 1.5~RTT, compared to 0.5~RTT when a packet is successfully transmitted or recovered on the lossy link. As we increase the coding rate and subsequently decrease the rate of retransmissions, the E2E delay approaches 0.5~RTT.

\section{Conclusion and Future Work }\label{sec:conclusion}
In this paper, we present \name, an in-network system to tame the impact of environmental packet loss events in hybrid wireless-fiber backbone networks. We provide a detailed analysis of \name's goodput gains, providing insights on choosing optimal parameters in practical settings. Our simulations show both the validity of our analysis and that \name significantly reduces the rate of E2E retransmissions, yielding higher goodput in the network and lower delay for flows subject to packet loss events. 
Future work will include more sophisticated coding strategies, as well as a programmable switch implementation.

\bibliographystyle{ACM-Reference-Format}
\bibliography{ref}

\end{document}